\documentclass[acmsmall]{acmart}

\setcopyright{cc}
\setcctype{by}
\acmDOI{10.1145/3728931}
\acmYear{2025}
\acmJournal{PACMSE}
\acmVolume{2}
\acmNumber{ISSTA}
\acmArticle{ISSTA056}
\acmMonth{7}
\received{2025-02-26}
\received[accepted]{2025-03-31}

\usepackage{listings}
\usepackage{bm}
\usepackage{xspace}
\usepackage{verbatim}
\usepackage{tcolorbox}
\usepackage{stfloats}
\usepackage{textcomp}
\usepackage{booktabs, tabularx}
\usepackage{rotating}
\usepackage{color, colortbl}
\usepackage{xcolor}
\usepackage{soul}
\soulregister\ref7
\soulregister\cite7
\usepackage{fancybox}
\usepackage{multirow}
\usepackage{enumitem}
\usepackage{soul}

\usepackage{url}

\usepackage[caption=false,font=footnotesize]{subfig}
\usepackage{balance}
\usepackage{footnote}

\usepackage{array}

\usepackage{adjustbox}
\usepackage{threeparttable}
\usepackage{multirow}
\usepackage[normalem]{ulem}
\useunder{\uline}{\ul}{}

\definecolor{dkgreen}{rgb}{0,0.6,0}
\definecolor{gray}{rgb}{0.5,0.5,0.5}
\definecolor{mauve}{rgb}{0.58,0,0.82}
\definecolor{dkred}{rgb}{0.6, 0.1, 0}
\definecolor{dkblue}{rgb}{0, 0, 0.6}

\newcommand{\phead}[1]{\vspace{1mm} \noindent {\bf #1}}

\newcommand{\change}[1]{#1}

\newcommand{\tool}{ClearCRC\xspace}

\newcommand{\greybox}[2]{\begin{tcolorbox}[width=\linewidth,boxrule=0pt,top=1pt, bottom=1pt, left=1pt,right=1pt, colback=gray!20,colframe=gray!20]
{\textbf{#1}: }{#2}
\end{tcolorbox}}

\definecolor{mygreen}{rgb}{0,0.6,0}
\definecolor{mygray}{rgb}{0.5,0.5,0.5}
\definecolor{mymauve}{rgb}{0.58,0,0.82}

\DeclareCaptionFormat{mylst}{\hrule#1#2#3}
\usepackage{array}

\begin{document}

\title{Understanding Practitioners’ Expectations on Clear Code Review Comments}

\author{Junkai Chen}
\authornote{~Equal contribution.}
\orcid{0009-0000-9945-7729}
\affiliation{%
  \institution{The State Key Laboratory of Blockchain and Data Security, Zhejiang University}
  \city{Hangzhou}
  \country{China}
}
\email{junkaichen@zju.edu.cn}

\author{Zhenhao Li}
\authornotemark[1]
\orcid{0000-0002-4909-1535}
\affiliation{%
  \institution{York University}
  \city{Toronto}
  \country{Canada}
}
\email{lzhenhao@yorku.ca}

\author{Qiheng Mao}
\orcid{0000-0002-7259-1087}
\affiliation{%
  \institution{Zhejiang University}
  \city{Hangzhou}
  \country{China}
}
\email{maoqiheng@zju.edu.cn}

\author{Xing Hu}
\authornote{~Corresponding author.}
\orcid{0000-0003-0093-3292}
\affiliation{%
  \institution{The State Key Laboratory of Blockchain and Data Security, Zhejiang University}
  \city{Hangzhou}
  \country{China}
}
\email{xinghu@zju.edu.cn}

\author{Kui Liu}
\orcid{0000-0003-0145-615X}
\affiliation{%
  \institution{Huawei}
  \city{Hangzhou}
  \country{China}
}
\email{brucekuiliu@gmail.com}

\author{Xin Xia}
\orcid{0000-0002-6302-3256}
\affiliation{%
  \institution{Zhejiang University}
  \city{Hangzhou}
  \country{China}
}
\email{xin.xia@acm.org}

\renewcommand{\shortauthors}{Chen et al.}

\begin{abstract}
\definecolor{deepred}{rgb}{0.6, 0, 0} 
\textcolor{deepred}{\textbf{WARNING}: This paper contains potentially offensive and harmful content.}

\noindent The code review comment (CRC) is pivotal in the process of modern code review. It provides reviewers with the opportunity to identify potential bugs, offer constructive feedback, and suggest improvements. Clear and concise code review comments (CRCs) facilitate the communication between developers and are crucial to the correct understanding of the identified issues and proposed solutions. Despite the importance of CRCs' clarity, there is still a lack of guidelines on what constitutes a good clarity and how to evaluate it. In this paper, we conduct a comprehensive study on understanding and evaluating the clarity of CRCs. We first derive a set of attributes related to the clarity of CRCs, namely \textsf{RIE} attributes (i.e., \textit{Relevance}, \textit{Informativeness}, and \textit{Expression}), as well as their corresponding evaluation criteria based on our literature review and survey with practitioners. We then investigate the clarity of CRCs in open-source projects written in nine programming languages and find that a large portion (i.e., 28.8\%) of the CRCs lack the clarity in at least one of the attributes. Finally, we explore the potential of automatically evaluating the clarity of CRCs by proposing \tool. Experimental results show that \tool with pre-trained language models is promising for effective evaluation of the clarity of CRCs, achieving a balanced accuracy up to 73.04\% and a F-1 score up to 94.61\%.

\end{abstract}




\maketitle

\section{Introduction}
\label{sec:intro}

Code review is the process of systematic examinations on software source code performed by third-party developers~\cite{6148202, mcintosh2014impact}. The primary goals of code review include identifying potential issues, seeking areas for improvement, and transferring knowledge~\cite{sadowski2018modern,yang2024survey,beller2014modern,10.1145/3239235.3239525, kononenko2015investigating}. It has been widely integrated into the software development life cycle in both open-source and industrial projects to help the assurance of software quality.

\begin{figure}
 \centering
\includegraphics[width=0.75\linewidth]{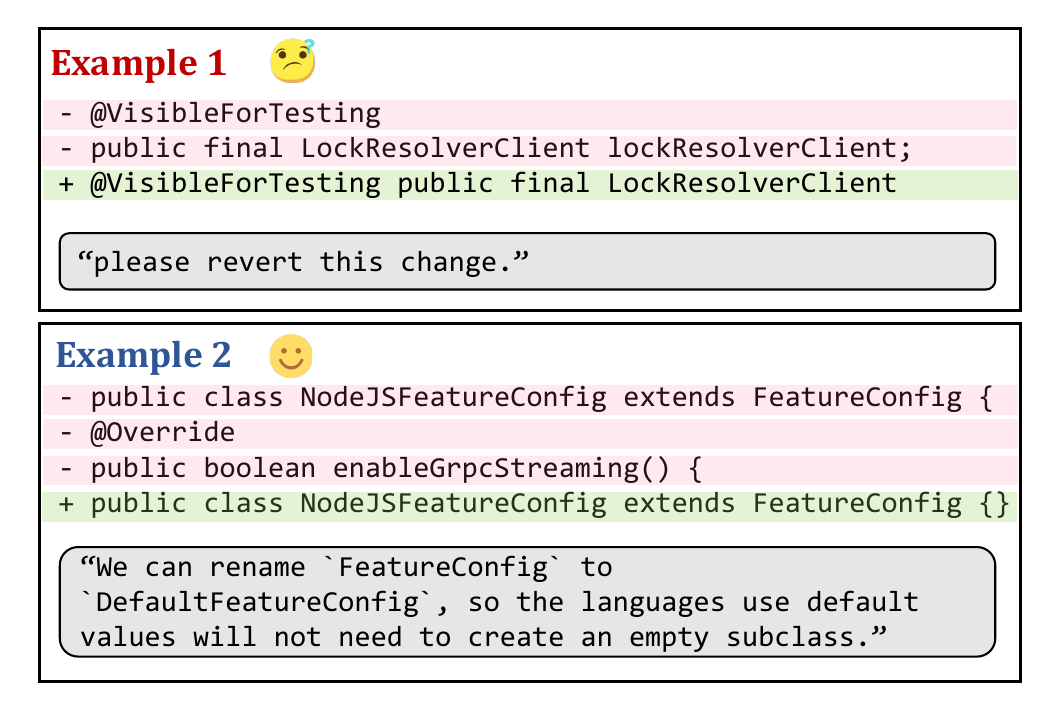}
 \caption{Examples of code changes and their code review comments (CRCs).}
 \label{fig:intro_example}
 \vspace{-0.3cm}
 \end{figure}

\vspace{-0.0cm}
A code review comment (CRC) is a specific piece of feedback provided by a reviewer during the code review process.
Clear and concise code review comments (CRCs) are crucial for ensuring that the feedback is readable and actionable, and further contributing to the overall quality of the software. On the contrary, CRCs that lack of sufficient clarity may lead to confusion, misunderstandings, and misinterpretations amongst the collaborating developers.
Figure~\ref{fig:intro_example} presents two examples of code changes and their corresponding code review comments. In Example 1, the reviewer comments {\em ``please revert this change''}.  However, no rationale or reason behind this comment is provided. Developers may not understand why this change needs to be reverted. In Example 2, the reviewer suggests renaming a parent class and also explains the reason of such suggestion. Moreover, this comment is written in a more friendly tone (i.e., {\em ``We can ...''}). Although both of these two examples provide a suggestion to modify the code, their effectiveness in conveying reviewer's idea may considerably vary.

Prior studies provide preliminary insights on revealing the quality of CRCs. 
For example, usefulness~\cite{bosu2015characteristics,7962371,hasan2021using} focuses on whether the CRCs can trigger subsequent code changes or if the reply to CRCs has a positive sentiment (e.g., {\em ``LGTM''}).
Yang et al.~\cite{yang2023evacrc} proposed four attributes to evaluate the quality of CRCs. Such attributes focus more on the purpose of CRCs (e.g., evaluation, suggestion, and question). However, these studies either indirectly evaluate the quality of CRCs using the information after the completion of the review, or evaluate the CRCs based on whether it has elements related to the purpose of this comment.
Therefore, a systematic understanding and characterizing of how a CRC can clearly and concisely foster the communication among developers (i.e., clarity of CRCs) is still in mystery and an on-going challenge.

In this paper, we conduct a comprehensive study to uncover the clarity of CRCs by following a multi-phased investigation: 1) we understand the characteristics and evaluation criteria of CRCs' clarity through a systematic literature review, a preliminary review with industrial professionals, and an online questionnaire survey with practitioners; 2) we examine the clarity of CRCs in open-source projects by conducting a manual investigation on sampled datasets; 3) we seek to automatically evaluate the clarity of CRCs by proposing an automated framework.

Particularly, we study the clarity of code review comments by answering three research questions:

\phead{RQ1: What attributes are relevant to the clarity of CRC?}
Based on the analysis on our literature review and 103 survey responses from practitioners, we derive our {\sf RIE} attributes for the clarity of CRCs (i.e., {\em Relevance}, {\em Informativeness}, and {\em Expression}) and their corresponding evaluation criteria. More than 75\% of the participants consider that these attributes are important to the clarity of CRCs.

\phead{RQ2: How is the clarity of code review comments in open-source projects?}
We manually investigate the clarity of code review comments in open-source projects using the datasets sampled from the work of Li et al.~\cite{codereviewer}. We find that 28.8\% of the CRCs in our study datasets are insufficient in at least one of the three attributes of CRCs' clarity. Among these attributes, {\em Informativeness} has the most considerable insufficiency.

\phead{RQ3: Can we automatically evaluate the clarity of code review comments?}
We propose \tool, an automated framework for the evaluation of CRCs' clarity based on the {\sf RIE} attributes. We compare the results of \tool with three sets of backbone models: 1) training deep learning and machine learning models; 2) fine-tuning pre-trained language models; and 3) prompting large language models (LLMs). Our results show that \tool is effective in evaluating each of the {\sf RIE} attributes, with an average balanced accuracy of 73.04\% using pre-trained language models.

We summarize the contributions of this paper as follows:

\begin{itemize}
    \item We derive {\sf RIE} attributes and their corresponding evaluation criteria for the clarity of CRCs by analyzing the results of our literature review and 103 survey responses from practitioners around the world.
    \item We find that a large portion of the CRCs in open source projects actually lack of sufficient clarity. We also publicly share our manually labelled data in the replication package~\cite{replication_package} for future studies. 
    \item We propose \tool, an automated framework for the evaluation of CRCs' clarity using various backbone models such as machine learning, deep learning, pre-trained language models, and large language models. \tool achieves promising results in our evaluation, especially using pre-trained language models.

\end{itemize}

Overall, the findings of our studies may be used as actionable guidelines for evaluating and writing clear CRCs, as well as for curating high-quality data to improve the automated generation techniques of CRCs.

\phead{Paper Organization.}
 Section~\ref{sec:related} summarizes the related work. 
 Section~\ref{sec:methodology} presents the methodology of our study. 
 Section~\ref{sec:results} discusses the results of our research questions.
 Section~\ref{sec:discussion} discusses the implications of our study. 
  Section~\ref{sec:threats} discusses the threats to validity. 
 Section~\ref{sec:conclusion} concludes the paper.
\section{Related Work}
\label{sec:related}

In this section, we summarize the related work in two aspects: studying the quality of code review comments and automated generation of code review comments.

\phead{Quality of Code Review Comments.}
Kerzazi et al.~\cite{ASRI201937} found that sentiment conveyed within comments can significantly impact the outcome of the review process. 
Kononenko et al.~\cite{7886977} suggested that the review quality is mainly associated with the thoroughness of the feedback, the reviewer's familiarity with the code, and the perceived quality of the code itself.
Rahman et al.~\cite{7962371} presented a comparative analysis of useful versus non-useful review comments, distinguishing them through their textual attributes and the reviewers' expertise. Comments were classified as useful or non-useful depending on their capacity to instigate changes.
Chouchen et al.~\cite{chouchen2021anti} synthesized negative examples of code reviews that degraded software quality, categorizing erroneous practices into five patterns: Confused reviewers, Divergent reviewers, Low review participation, Shallow review, and Toxic review.
Ram et al.~\cite{ram2018makes} focused on the issue of code change reviewability, which is closely related to the quality of reviews.
Ebert et al.~\cite{ebert2019confusion, ebert2021exploratory} identified confusion as a significant detriment to the quality of code reviews and offered recommendations for addressing issues of confusion.
Bosu et al.~\cite{bosu2015characteristics} emphasized that the usefulness of code review lies in its ability to assist developers in avoiding defects, adhering to team conventions, and resolving issues efficiently and reasonably.
Ferreira et al.~\cite{ferreira2021shut} highlighted the prevalence of uncivil behavior during the code review process, noting that discourteous comments can hinder project communication and discussion, ultimately slowing down development progress.
Pascarella et al.~\cite{pascarella2018information} examined the essential information required by reviewers in code reviews, including the suitability of an alternative solution, correct understanding, rationale, code context, etc.
Yang et al.~\cite{yang2023evacrc} introduced four attributes for evaluating the quality of CRC: questions, suggestions, evaluations, and emotion, advocating for an assessment of the quality. 

Prior studies provide insights on the quality of CRCs from different perspectives. These studies generally emphasize the importance of {\em ``clear''} CRCs. However, the  understanding and characterizing on what are {\em ``clear''} CRCs are still limited. Therefore, in this paper, we conduct a comprehensive study to uncover and demystify what are the characteristics of a {\em ``clear''} CRC.

\phead{Automated Generation of Code Review Comments.}
The automated generation of code review comments has been widely studied in recent years~\cite{siow2020core,tufano2021towards,lu2023llama,li2022automating}, aiming to streamline this critical yet often labor-intensive activity in the software engineering life cycle. Efforts in this domain can be categorized into three main types of approaches: traditional rule-based methods, deep learning techniques, and Large Language Models (LLMs) based techniques.
Early automation efforts in code review were aimed at identifying code violations and defects utilizing traditional rule-based static analysis tools~\cite{balachandran2013reducing}. While providing a base for automation, they lacked the flexibility to adapt to the nuanced and evolving nature of software development practices. 
The emergence of deep learning has significantly enhanced the capability to automate code reviews, offering a nuanced understanding and interpretation of code changes. Techniques leveraging LSTM~\cite{gupta2018intelligent}, and Transformers~\cite{li2022automating, tufano2022using} have been pivotal in predicting review necessities and generating context-specific feedback. 
Li et al.~\cite{li2022automating} proposed a pre-trained model based on the Text-To-Text-Transfer Transformer (T5) model~\cite{raffel2020exploring}, specifically tailored for the code review process across three different code review tasks including the generation of CRCs.
LLaMA-Reviewer~\cite{lu2023llama} utilized the LLaMA model and parameter-efficient fine-tuning to automate code review comments generation, achieving performance on par with existing code-review-focused models using fewer resources. 

Prior studies commonly incorporated the CRCs data directly, without a curation or quality selection process. Given this scenario, existing CRCs generation techniques might inadvertently learn from CRCs data lacking clarity, resulting in perplexing outcomes. Our study can complement the research of automated CRCs generation to curate the training data, and further improve the quality of the generated CRCs.

\section{Methodology}
\label{sec:methodology}

\begin{figure*}
 \centering
\includegraphics[width=0.99\linewidth]{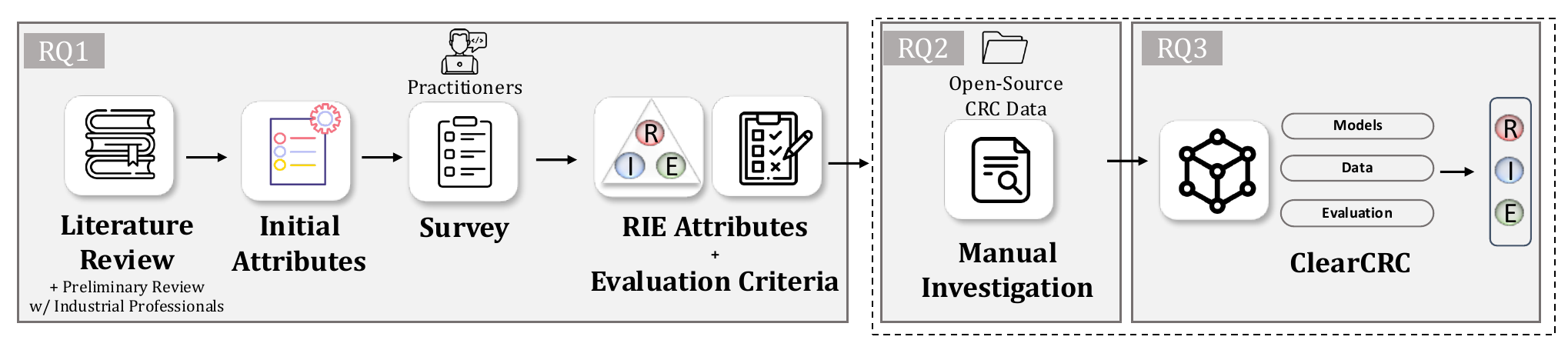}
 \caption{Overview of our study.}
 \label{fig:overall}
\vspace{-0.3cm}
 \end{figure*}

Figure~\ref{fig:overall} presents an overview of our study. To address the three research questions proposed in the Introduction, we conduct a comprehensive study that involves three phases.
\textbf{{\em Phase 1:}} We first derive an initial list of attributes and evaluation criteria concerning the clarity of CRCs through literature review and a preliminary review with industrial professionals. We then survey with practitioners for their perspectives to refine the attributes and evaluation criteria.
\textbf{{\em Phase 2:}} We manually investigate the clarity of CRCs in open-source projects using the clarity attributes and evaluation criteria derived in the prior phase.
\textbf{{\em Phase 3:}} We propose \tool, an automated framework that seeks to evaluate the clarity of CRCs based on the attributes derived from our literature review and practitioners' feedback.

\subsection{Characterizing the Clarity of CRCs}

We characterize the attributes related to the clarity of CRCs by combining 1) a systematic literature review followed by a preliminary review with industrial professionals, and 2) the analysis of our online survey with practitioners.

\subsubsection{\textbf{Literature Review}} We analyze existing studies on the quality of CRCs and derive an initial set of attributes related to the clarity of CRCs.

\phead{Literature Collection.}
We first gather the papers related to code review by utilizing the list of papers provided by prior literature reviews~\cite{code_review_survey_davila,code_review_survey_wang}. These two literature reviews summarized 139 and 112 prior works on code review published from 2005 to 2019 and 2011 to 2019, respectively.
We then follow the strategy of paper collection in these literature reviews to collect papers published after 2019 and collect 70 papers in this process.

\phead{Data Analysis.}
We first read the titles and abstracts of all the collected papers and filter papers that are not related to the studies on CRCs. The topics of such filtered papers include studying the code change, authors, reviewers, pull requests, and commit messages. After this process, we have a list of 47 papers that are related to CRCs for further analysis. 
Two authors of this paper then independently read these papers and generate an initial set of codes related to the attribute of CRCs' clarity. 
The authors then perform open card sorting~\cite{open_card_sorting} on the generated codes to analyze the codes and sort the generated codes into potential themes that indicate the attributes related to the clarity of CRCs. 
Particularly, author 1 generates six initial codes: \textit{Confused Reviews, Toxic Reviews, Relevant Reviews, Readable Reviews, Shallow Reviews, and Informative Reviews}. Author 2 generates five initial codes: \textit{Readability, Sentiment, Relevance, Information, and Toxicity}.
The two authors then engage in thorough discussions to refine the attribute set. Both authors collaboratively re-evaluate the attributes and reconcile differences by focusing on semantic consistency and conceptual clarity. Throughout this process, attributes that exhibit overlap or ambiguity are merged or redefined. For example, \textit{Readability and Toxicity} are merged into \textit{Expression}, and the meaning of \textit{Shallow Reviews} can be represented by \textit{Informativeness}.
Eventually, we derive three attributes that are related to the clarity of CRCs, including {\bf {\em Relevance}}, {\bf {\em Informativeness}}, and {\bf {\em Expression}}.  The authors then discuss and generate an initial definition for each attribute.

\phead{Preliminary Review.}
To preliminarily verify the attributes derived from our literature review, we conduct a group interview with 11 industrial practitioners to obtain their feedback. The participants are full-time engineers at Company X, which is a world-leading IT company. All of the participants have at least three years of experience in code review and software development. The duration of this group interview is around 1 hour. We follow a three-step process to conduct the group interview: 1) We ask the participants to freely talk about their expectations on the clarity of CRCs without the knowledge of our derived attributes; 2) We present our derived attributes to the participants; 3) We discuss with the participants for whether the attributes can reflect their expectations on the clarity of CRCs.
Eventually, we find that most of the points initially proposed by the participants can be reflected by our derived attributes. The remaining non-reflected points are related to the process of code review rather than the code review comments, including \textit{deciding proper reviewers} and \textit{prompt response time}. 
At the end, the participants are thanked and briefly informed about the next plans.
The participants also provide suggestions regarding the design of surveys. We take their suggestions into consideration while designing our survey in the next step.

\subsubsection{\textbf{Questionnaire Survey with Practitioners}} We conduct an online questionnaire survey with practitioners for their perspectives on the clarity of CRCs and further refine the attributes and derive the evaluation criteria.

 \phead{Survey Design.} The questions in our survey are divided into three parts:

\begin{itemize}
    \item[P1] We ask the participants for their basic information (e.g., country or region of residence, primary job role, and years of experience in the primary job role) and if they have experience in code review.
    \item[P2] For each attribute of the clarity, we ask the participants for ``{\em To what extent do you think that the aspect of "\{attribute\}" is important to the clarity of code review comments?}''~where \textit{\{attribute\}} is replaced with a specific one. The participants then choose from ``Very important'', ``Important'', ``Neutral'', ``Unimportant'', and ``Very unimportant''. If ``Very important'' or ``important'' is chosen, we further ask the participants for the details of how they will evaluate the corresponding attribute (e.g., what factors and detailed information they may focus on). We will derive the evaluation criteria based on their comments. If other options are chosen, we ask the participant for the potential reasons.
    At the end of this part, we further ask the participant for ``{\em Do you have some ideas about other important aspects that contribute to the clarity of code review comments?}''.
    \item[P3] We ask the participants for questions on automated code review comment generation, such as their experience in using such tools, and their perspectives on the clarity of the generated CRCs.

\end{itemize}

 \phead{Survey Implementation.}
We implement the survey following the design discussed above using Microsoft Forms~\cite{microsoft_forms}.
We conduct a pilot survey with three practitioners to collect their feedback on the design of our survey. All the practitioners in the pilot survey have experience in writing and reviewing CRCs. 
The pilot participants provide suggestions regarding the clarification of instructions and the consistency of some terms.
We make modifications according to their feedback and have a final version of the survey, which is an anonymous questionnaire and can be accessed using the link provided in our email sent to the participants.
 A sample of the complete survey is available in our replication package~\cite{replication_package}.

 \phead{Participants.}
 To invite participants from diverse backgrounds, we reach out to industrial and academic professionals residing in various countries or regions, across five continents around the world.
 Eventually, we receive 112 responses in total. We then filter 9 responses of which indicate as having no experience in code review, resulting in a total of 103 remaining responses for further analysis.

\noindent \underline{Demographics.}
Table~\ref{table:demographics} shows the statistical information of the participants.
The participants reside in 37 countries or regions across five continents, including 49 participants in Europe, 22 participants in Asia, 24 participants in North America, 4 participants in South America, and 4 participants in Oceania. 
A majority of the participants have an occupation of industrial/freelance professional (76.7\%) and primary job role as development (81.6\%). A large percentage of the participants (68.9\%) have at least five years of experience in their primary job role. In short, our survey participants reside in various countries or regions all over the world, and most of them are industrial or freelance professionals with more than 5 years of experience in software development.

\begin{table*}
    \caption{Statistics of our survey participants. }
    \centering
    
    \tabcolsep=10pt
\resizebox{\linewidth}{!} {
    \begin{tabular}{lr|lr|lr|lr}
        \hline
\rowcolor[HTML]{EFEFEF}    \multicolumn{2}{c|}{\textbf{Residency}}  &   \multicolumn{2}{c|}{\textbf{Occupation}}  &  \multicolumn{2}{c|}{\textbf{Primary Job Role}}  &  \multicolumn{2}{c}{\textbf{Experience}}   \\
        \hline
        Asia   & 22    & Academic or industrial researcher   &8    &    Development   & 84    &   [0-2]   & 10   \\
        Europe   & 49     &  Graduate or undergraduate student   & 12     &   Software Project Management   & 5    &   [3-5]   & 22   \\
        North America   & 24     &  Industrial or freelance professional   & 79     &   Testing   & 2    &   [6-9]   & 19   \\
        Oceania  & 4     &  Other   & 4    &    Research   & 9    &   [10+]   & 52   \\
        South America   & 4     &  -   & -    &    Other   & 3    &   -  & -   \\
        \hline
    \end{tabular}
    }
    \vspace{-0.0cm}
    \label{table:demographics}
\end{table*}

\phead{Data Analysis.}
The data we obtain from the survey consists of option data from multiple-choice questions and natural language response data from open-ended questions.
(1) For the questions of multiple choices, we compute the percentage of each option, e.g., the percentage of survey participants who think "Relevance" is "Very Important" to the clarity of CRCs shown in Fig.~\ref{fig:attributes}.
(2) For the open questions (e.g., details for evaluating the attributes), 
we generate codes from the answers and perform open card sorting~\cite{open_card_sorting} to analyze the thematic similarity.
Specifically, to derive the evaluation criteria for each attribute, we first extract and record criteria of all the responses, and then sort and categorize them into concise and explicit descriptions like "Proper syntax and grammar". For example, a response from one survey participant for \textit{Informativeness} said ``\textit{Explain why, with specific reference to the change.}'' will finally lead to two criteria including ``I.E2: Provide reasons or context information'' and ``I.O2: Provide reference information'' (See Section~\ref{sec:results}).
When a consensus on these criteria for each attribute is reached, we first filter evaluation criteria mentioned fewer than five times (i.e., 1-4 times), and then select evaluation criteria which are mentioned 15+ times as {\em essential} evaluation criteria and the remaining ones as {\em optional} evaluation criteria. We take them as references for manual investigation.
Detailed results will be presented in Section~\ref{sec:results} (RQ1).

\subsection{Investigating the Clarity of CRCs in Open-Source Projects}

In this phase, we manually investigate the clarity of CRCs in open-source projects using the attributes and evaluation criteria of clarity derived in the prior phase.

\phead{Data Preparation.}
We use the benchmark dataset proposed by Li et al.~\cite{codereviewer} to conduct manual investigation.
The dataset contains pairs of diff hunk and CRC written in nine programming languages.
Specifically, we randomly sample a set of data from its validation dataset for each programming language. We do not sample from its training dataset because the data in different programming language is combined together and can not be distinguished.
For each programming language, we randomly sample a set of data based on 95\% confidence level and 5\% confidence interval~\cite{sample}. 
Table~\ref{table:rq2_dataset} presents the details of our sampled datasets.
In total, we randomly sample 2,438 pairs of diff hunk and CRC. The sample size of each programming language varies from 216 for C to 339 for Golang.

\phead{Manual Investigation.}
Two authors of this paper first carefully read the attributes and evaluation criteria derived in the previous phase, and discuss until the two authors have a clear and consistent understanding on the details. For each sampled data, the two authors then independently 
examine the CRC and its corresponding code change to label if it meets the evaluation criteria for each attribute.
When the process of labelling is completed, the two authors compare their results and discuss each disagreement until reaching a consensus. We have a Cohen's Kappa~\cite{kappa} value of 0.87 in this process, which indicates a substantial agreement.
We will discuss the results of our manual investigate in Section~\ref{sec:results} (RQ2).

\begin{table}
    \caption{An overview of our studied open-source dataset. }

    \centering
    
    \tabcolsep=17pt
    \begin{tabular}{l|rr}
        \hline
\rowcolor[HTML]{EFEFEF}  \textbf{Language}  &   \textbf{Original \#}   &\textbf{Sampled \#}   \\
        \hline
        \textbf{C}   & 492   & 216\\
        \textbf{C++}   & 736   & 253\\   
        \textbf{C\#}   & 682   & 246\\
        \textbf{Golang}   & 2,826   & 339\\
        \textbf{Java}   & 1,636   & 312\\
        \textbf{JavaScript}   & 1,035   & 281\\
        \textbf{PHP}  & 443   & 206\\
        \textbf{Python}  & 1,420   & 303\\
        \textbf{Ruby}  & 1,049   & 282\\

        \hline
        \textit{Total}  & 10,319  &2,438\\
        \hline
    \end{tabular}

    \vspace{-0.0cm}

    \label{table:rq2_dataset}

\end{table}

\subsection{\tool: Automatically Evaluating the Clarity of CRCs}
In this phase,  we propose \tool, an automated framework that aims at the evaluation of the clarity of CRCs, based on the {\sf RIE} attributes derived from our literature review and practitioners' feedback. We adopt different sets of backbone models in our framework to empirically study their effectiveness in automatically evaluating the clarity of CRCs.

\phead{Models.} We use three sets of backbone models, including deep learning and machine learning models, pre-trained language models (e.g., CodeBERT~\cite{codebert} and CodeReviewer~\cite{codereviewer}), and large language models (e.g., Llama~\cite{llama} and CodeLlama~\cite{codellama}).

\noindent \underline{Model Set 1: Deep Learning and Machine Learning Models.}
Prior studies on classifying good commit messages~\cite{good_commit_message} and log messages~\cite{log_message_ase2023} indicate that Bi-LSTM and Random Forest are effective in such classifications. Following these studies, we use Bi-LSTM~\cite{bi-lstm} and Random Forest~\cite{random_forests} as the deep learning and machine learning based backbones to perform the evaluation of CRCs' clarity.

\noindent \underline{Model Set 2: Pre-trained Language Models.}
For pre-trained language models, we use CodeBERT~\cite{codebert} and CodeReviewer~\cite{li2022automating} as our subject techniques.
CodeBERT is a bimodal pre-trained language model for programming languages and natural languages with the same model architecture as RoBERTa-base~\cite{liu2019roberta}. It has been widely used by prior studies for classification tasks and presents a promising balance between performance and cost of computing resources~\cite{zeng2024colare,turzo2023towards,zhang2023vulnerability}.
CodeReviewer is a pre-trained model specialized for the automation of code review activities. They proposed pre-training tasks designed for code review like code diff denoising, and then pre-trained the CodeT5~\cite{wang2021codet5} model on a large-scale code review dataset. CodeReviewer shows competitive performance on code review tasks such as code refinement.

\noindent \underline{Model Set 3: Large Language Models.}
LLMs have demonstrated promising results in various software engineering tasks~\cite{liu2024large,chen2024code,chen2025reasoning}, which brings opportunities and challenges for using LLMs as evaluators~\cite{li2024split,chen2024nlperturbator}.
For LLM baselines, we use {\em Llama3-70B-Instruct}~\cite{llama} and {\em CodeLlama-34B-Instruct}~\cite{codellama}. We choose them since Llama series models show great performance among different LLMs~\cite{llama,codellama}, and they are very popular in research related to code review~\cite{lu2023llama, yu2024fine, zhao2023right}.
Additionally, we also attempt to include a code-review-specialized LLM named LLaMA-Reviewer~\cite{lu2023llama}. However, since LLaMA-Reviewer is tailored to specific downstream tasks and does not generalize well to our setting (i.e., it tends to generate invalid outputs when prompted), we finally decide to exclude it.

\phead{Data.}
Here we introduce the datasets we use as well as their augmentation and pre-processing.

\noindent \underline{Datasets and Augmentation.}
We utilize the manually labelled datasets in the prior phase to conduct the study, which consists of 2,438 pairs of code change and CRC in total. We randomly split the datasets into 80\% training, 10\% validation, and 10\% testing. 
As shown in Table~\ref{table:RQ2}, the distribution of negative and positive instances is imbalanced in the dataset. To mitigate such impact, for each experiment, we perform up-sampling on the corresponding attribute. 
Specifically, we randomly repeat the negative instances of the experimented attribute in the training dataset to have the same amount as the positive instances. Note that we only augment data in the training set and ensure the testing set is consistent for all backbone models.

\begin{figure}
 \centering
\includegraphics[width=1.00\linewidth]{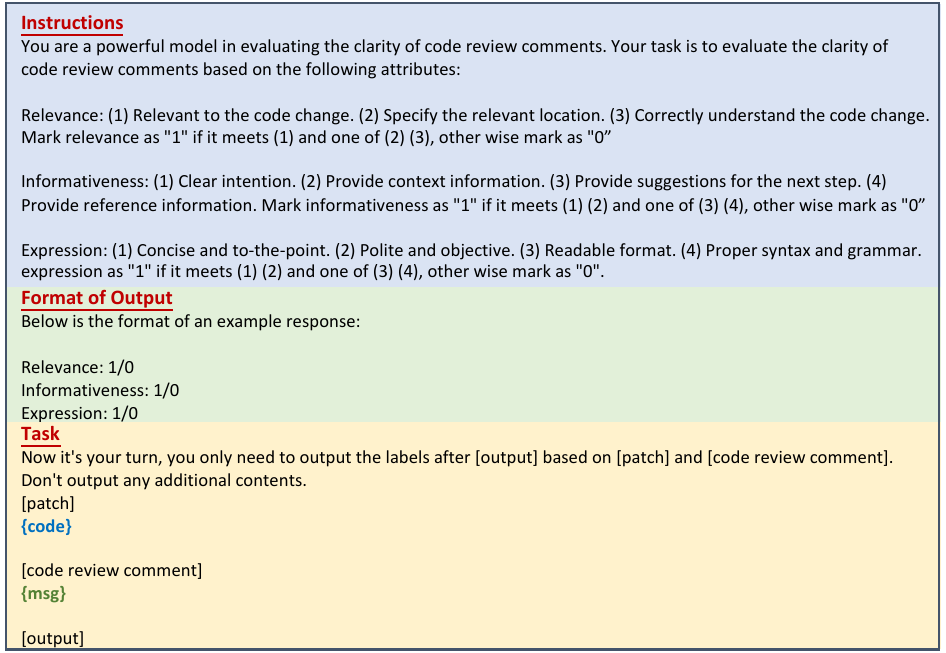}
 \caption{Prompt Template for using LLMs to evaluate the clarity of CRCs.}
 \label{fig:prompt}
\vspace{-0.3cm}
 \end{figure}

\noindent \underline{Processing.}
We analyze the raw input data including pairs of code change and CRC, process and combine the data with code change and CRC to feed into the model. We remove the lines of code that are unrelated to the code changes (e.g., the surrounding code of code changes). 
For models in set 1\&2, we replace the {\em ``-''} and {\em ``+''} mark at the start of each line with {\em ``[DELETE]''} and {\em ``[ADD]''}~in the code change, respectively.  We then concatenate the CRC and the processed code change together, and attach a {\em ``[SEP]''} token between them.
For large language models in set 3, We embed the information of code change and CRC into the prompt and inference the models to obtain the results returned by the models and further evaluate the clarity of each attribute.
For the prompt of using LLMs, we follow Prompt Engineering Guide to design the prompts~\cite{prompt_guide}. As shown in Figure~\ref{fig:prompt}, we first provide an instruction of the task, the attributes, and evaluation criteria. We then inform the models of how to use the evaluation criteria (i.e., meet all of the essential ones and at least one of the optional ones) and the expected template of output. Finally, we attach the actual data (i.e., diff hunk and CRC) and have the model start its evaluation.

\phead{Evaluation.} We introduce the evaluation details in our empirical study. 

\noindent \underline{Metrics.}
We use four metrics to evaluate the results of \tool and the baselines: 1) balanced accuracy, 2) precision, 3) recall, and 4) F-1 score. 
Balanced accuracy is computed based on the average of true positive rate and true negative rate. A higher balanced accuracy indicates a better capability in identifying both positive and negative instances. The balanced accuracy of random guess in binary classification is close to 50.0\%~\cite{wheretolog_ASE20}.
It is widely used by prior work to evaluate the performance of binary classification, especially on imbalanced data~\cite{wheretolog_ASE20, Zhu:2015:LLH:2818754.2818807}.
For the calculation of precision, recall, and F-1 score which focus on the classification performance of {\em positive} instances, we consider CRCs that meet the evaluation criteria as {\em positive} instances, and otherwise as {\em negative} instances.

\noindent \underline{K-Fold Cross Validation.}
We utilize 5-fold cross validation to mitigate the impact of randomness, where 5 is a commonly used K value in prior studies involving k-fold cross validation~\cite{karal2020performance_5fold,ghorbani2020comparing_5fold,nti2021performance_kfold}. We randomly split the dataset into five subsets. The validation has five rounds in total. For each round of validation, we use one subset (i.e., 20\%) for validation and testing (i.e., half for validation and half for testing), and the remaining four subsets (i.e., 80\%) for training. We ensure that the dataset for each fold is identical for all models to perform a fair comparison.

\phead{Implementation Details.} (1) For models in set 1, we use PyTorch~\cite{pytorch} to implement Bi-LSTM and use Scikit-learn~\cite{sklearn} to implement Random Forests, respectively. We follow prior studies~\cite{good_commit_message,log_message_ase2023} to set the hyperparameters of the networks and the training processes.
(2) For pre-trained models in set 2, we access the models through the official checkpoints released on HuggingFace~\cite{huggging_face}. As to hyperparameters like batch size and learning rate, we tune them according to the official replication package and the volume of our datasets. During the training stage, we set the number of training epochs to 10 and perform the strategy of early stopping (n=3) on all models to limit the training consumption and save the best-performing model on the validation dataset for further testing. We adopt the AdamW~\cite{loshchilov2017decoupled} optimizer and linear scheduler.
(3) For large language models, we download their official checkpoints from HuggingFace. We set the maximum number of generated tokens to 32 which is appropriate for the output format, and keep the other parameters the same in the default configuration.
We publicly release the scripts, parameters, and datasets for further research~\cite{replication_package}.

\section{Results}
\label{sec:results}

In this section, we discuss the results of our RQs.

\subsection{\textbf{RQ1: Characterizing and Understanding the Clarity of CRCs}}
In this RQ, we discuss the results of our {\sf RIE} attributes and evaluation criteria of CRCs' clarity, derived from our literature review and practitioners' survey. Table~\ref{table:attributes} shows an overview of the attributes and their evaluation criteria. 
Below, for each attribute, we discuss its detailed evaluation criteria and our survey results.

\begin{figure}
 \centering
\includegraphics[width=0.8\linewidth]{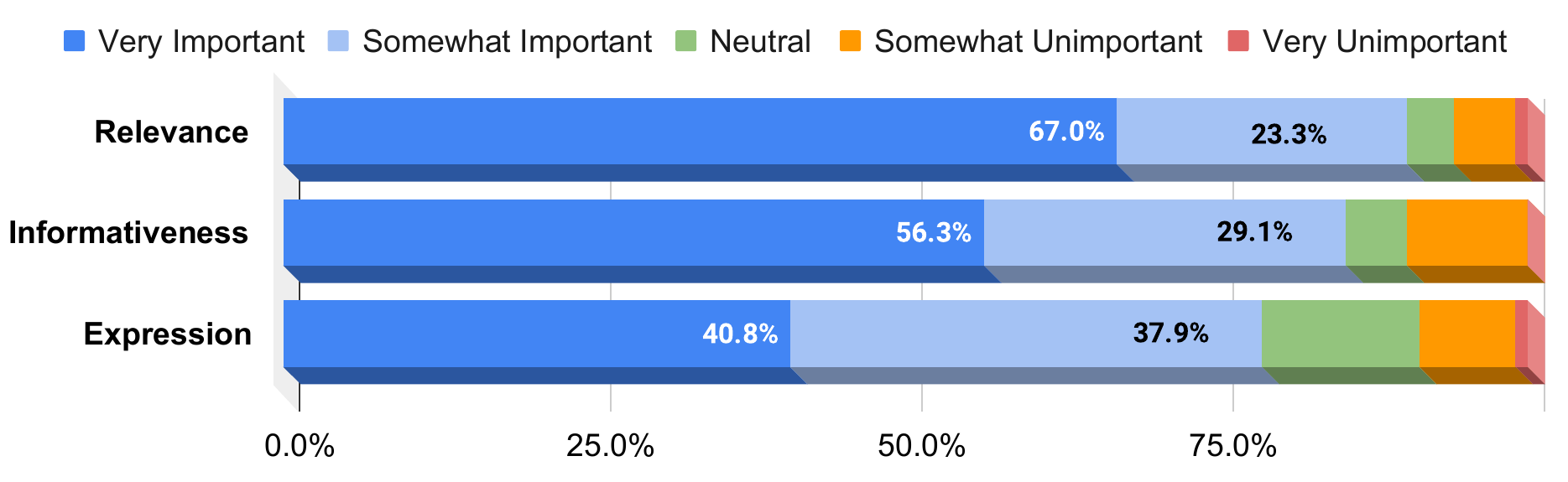}
 \caption{Survey participants' rating for each of the attributes (RQ1).}
 \label{fig:attributes}
 \vspace{-0.3cm}
 \end{figure}

\subsubsection{\textbf{Relevance}} If the code review comment is relevant to the code change.

\begin{figure}
 \centering
\includegraphics[width=0.8\linewidth]{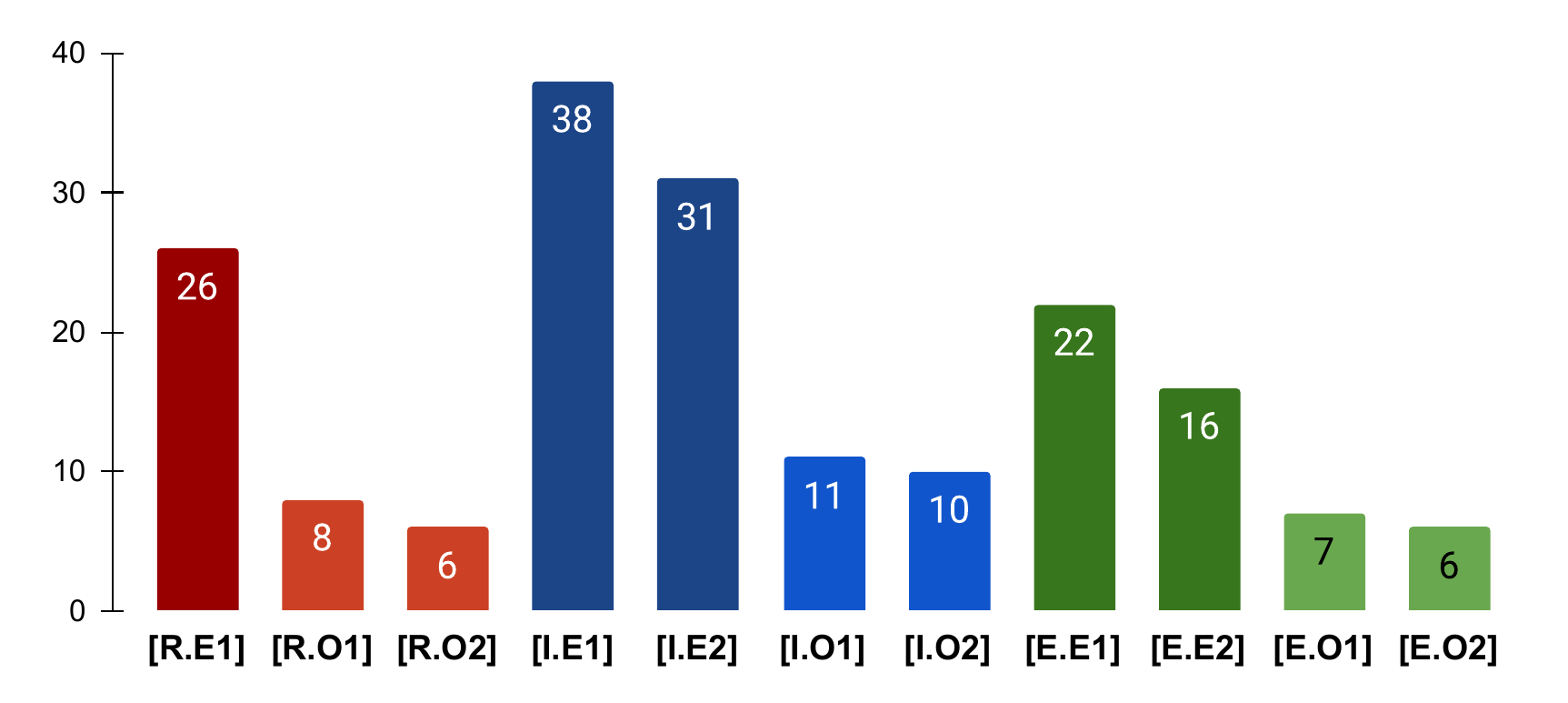}
 \caption{Frequency of each evaluation criteria mentioned in the survey.}
 \label{fig:evaluation_criteria}
 \vspace{-0.0cm}
 \end{figure}

\phead{Evaluation Criteria.} 
\change{After data analysis of open questions in our survey, we derive} one essential (i.e., R.E1) and two optional (i.e., R.O1 and R.O2) evaluation criteria corresponding to the attribute of relevance.  Figure~\ref{fig:evaluation_criteria} shows the frequency of evaluation criteria mentioned by our survey participants.

\begin{itemize}
\item[R.E1] \textbf{Relevant to the code change.} The CRC should be self-explanatory and relevant to the code change. It can be interpreted based on the current code change without a relying relevance to external information (e.g., other CRCs).
\item[R.O1] \textbf{Specify the relevant location.} The CRC specifies the particular position of the code which has the issues or concerns.
\item[R.O2] \textbf{Correctly understand the code change.} The CRC explicitly shows that the reviewer correctly understands the code change.
\end{itemize}

\phead{Discussion.}
Figure~\ref{fig:attributes} shows the percentage of each rate given by the survey participants regarding the importance of each attribute. Overall, most of the participants consider \textit{Relevance} is important to the clarity of CRCs, including 67.0\% as {\em very important} and 23.3\% as {\em somewhat important}. Below, we present the comments from our survey participants regarding their perspectives on the evaluation criteria of each attribute. We correspond such comments to the evaluation criteria discussed above, and the corresponding part is highlighted in bold.

\noindent \underline{R.E1:} {\em ``If the comment is made about the \textbf{intent or the change itself, not the state of the code in general}.''}

\noindent \underline{R.O1:} {\em ``Comments that do not pertain to the change as a whole, should refer directly to the \textbf{code elements} that should be modified in order for approval (e.g., \textbf{variable name, line of code)}.''}

\noindent \underline{R.O2:} {\em ``A relevant comment is one that is specific to the change and \textbf{shows a deep understanding of the code}.''}

\begin{table}
    \caption{An overview of the \textsf{RIE} attributes and their evaluation criteria derived from the survey. }
    %\scriptsize
    \centering
    \resizebox{\columnwidth}{!} {
    \tabcolsep=25pt
    \renewcommand\arraystretch{1.2}
    \begin{tabular}{c|c|l|l}

        %\toprule
        \hline
\rowcolor[HTML]{EFEFEF}    \textbf{Attribute}   & \textbf{ID}  & \textbf{Type}  & \textbf{Description} \\
%\midrule
\hline

\multirow{3}{*}{
\begin{tabular}[l]{@{}p{0.8cm}@{}} \textbf{Rel.}  \end{tabular}
}
    & \textbf{R.E1} &  Essential  & Relevant to the code change.       \\
    & \textbf{R.O1} & Optional    & Specify the relevant location.      \\
    & \textbf{R.O2}  & Optional     & Correctly understand the code change.             \\
        \hline

\multirow{4}{*}{
\begin{tabular}[l]{@{}p{0.8cm}@{}} \textbf{Info.}  \end{tabular}
}
    & \textbf{I.E1}   & Essential   & Clear intention.             \\
    & \textbf{I.E2}  & Essential     & Provide reason or context information.         \\
    & \textbf{I.O1}   & Optional    & Provide suggestions for the next step.         \\
    & \textbf{I.O2}   & Optional    & Provide reference information.          \\
        \hline

\multirow{4}{*}{
\begin{tabular}[l]{@{}p{0.8cm}@{}} \textbf{Exp.}  \end{tabular}
}
    & \textbf{E.E1}  & Essential    & Concise and to-the-point.        \\
    & \textbf{E.E2}   & Essential    & Polite and objective.       \\
    & \textbf{E.O1}   & Optional    & Readable format.         \\
    & \textbf{E.O2}  & Optional     & Proper syntax and grammar.      \\
        \hline

    \end{tabular}
    }

    \vspace{-0.0cm}

    \label{table:attributes}
\end{table}

\subsubsection{\textbf{Informativeness}} If the code review comment provides sufficient information.

\phead{Evaluation Criteria.} Similar to the process discussed in the evaluation criteria of \textit{Relevance}, there are 2 essential and 2 optional evaluation criteria corresponding to the attribute of \textit{Informativeness}.

\begin{itemize}
\item[I.E1] \textbf{Clear intention.} The CRC clearly specifies its intention (i.e., what is the further action needed) to make sure the CRC is actionable. The intention can include: 1) raising a question and asking for an answer; 2) identifying a problem that should be fixed; 3) providing suggestions that may be non-blocking and not urgent to take action.
\item[I.E2] \textbf{Provide reason or context information.} Based on the intention, provide context in the CRC. For example, (1) questioning: specifying what is the point of the question (e.g., not just {``{\em Why?}''}); (2) identifying issues: explaining what is the problem; (3) providing suggestions: the reason of such suggestions.
\item[I.O1] \textbf{Provide suggestions for the next step.} Try to provide suggestions for the next step if available.
\item[I.O2] \textbf{Provide reference information.} The CRC provides reference information that might be helpful to the target developer; such information may include the link to reference documents, guidelines, code, etc.
\end{itemize}

\phead{Discussion.}
As shown in Figure~\ref{fig:attributes}, over 85\% of the participants consider that \textit{Informativeness} is important to the clarity of CRCs. The remaining participants consider its importance as neutral or somewhat unimportant. However, the participants do not leave comments regarding the potential reasons. Below, we present the comments from our survey participants for their suggested evaluation criteria on \textit{Informativeness}. Their comments are corresponded to the evaluation criteria discussed above and the related part is marked in bold.

\noindent \underline{I.E1:} {\em ``It should be immediately obvious after reading the comment \textbf{what the commenter wants me to do}, why, and why their version is better than my version of the change.''} 
{\em ``One of the most important aspects to me is if the \textbf{intent of the comment is clear}.''}

\noindent \underline{I.E2:} {\em ``Including \textbf{reason or context} of why the code review comment is made.''}
{\em ``Whether the comment contains a \textbf{reasoning for why} the change is wrong and needs to be amended.''}

\noindent \underline{I.O1:} {\em ``if the comment is rejecting a change, it should at least include a \textbf{suggestion of an alternative approach}.''}

\noindent \underline{I.O2:} {\em ``\textbf{Pointers and references to the materials} and existing discussions in the wild are important.''}

\subsubsection{\textbf{Expression}} If the code review comment is readable, easy to understand, and friendly.

\phead{Evaluation Criteria.} There are 2 essential and 2 optional evaluation criteria corresponding to the attribute of \textit{Expression}.

\begin{itemize}
\item[E.E1] \textbf{Concise and to-the-point.} Describe the idea as precise and concise as possible to avoid vagueness, ambiguity, and incoherence.
\item[E.E2] \textbf{Polite and objective.} The CRC should express the idea in a polite manner, and focus on the code rather than the person.
\item[E.O1] \textbf{Readable format.} The CRC is written in a human readable format.
\item[E.O2] \textbf{Proper syntax and grammar.}  The CRC is written in a correct syntax and grammar, without typos or incomplete words.
\end{itemize}

\phead{Discussion.}
As shown in Figure~\ref{fig:attributes}, 78.7\% (i.e., 40.8\% {\em very important} + 37.9\% {\em somewhat important}) of the survey participants acknowledge the importance of \textit{Expression} to the CRC's clarity. There are 12.6\% of the participants consider its importance as {\em neutral} and 8.8\% as {\em unimportant}. For example, one participant that selects {\em neutral} comments {\em ``It dependes on what the expression be used for. A long but easy to understant expression is ok''}. One participant that selects {\em somewhat unimportant} comments {\em ``Really depends on the working relations between the developer and the reviewer. As long as they can understand each-other all is well''}.
Overall, \textit{Expression} has a relatively lower positive rate compared to the other two attributes, but still accounts for the majority of the participants. 

Below, we present the comments regarding the evaluation criteria suggested by our survey participants who acknowledge the importance of \textit{Expression}.

\noindent \underline{E.E1:} {\em ``When evaluating the expression of a code review comment, you're looking at how well the feedback is communicated, whether it is clear, \textbf{concise, and effectively conveys} the reviewer's thoughts''}

\noindent \underline{E.E2:} {\em ``Comments should have \textbf{friendly tone and comment on the code, not the person}.''}

\noindent \underline{E.O1:} {\em ``\textbf{Formatting around non-english or code snippets}. (i.e. backticks ``). This helps improve overall clarity.''}

\noindent \underline{E.O2:} {\em ``Comment should be plain human readable sentences, because PR author and other reviewers are humans. \textbf{Proper syntax and grammar, absence of typos} are important as well.''}

\subsubsection{\textbf{Practitioners' Feedback on Additional Attributes.}}
Apart from the three attributes, we also ask the participants for additional aspects or attributes that may contribute to the clarity of CRCs.
In total, we receive 48 responses to this question. After removing three responses that are {\em ``N/A''} or {\em ``None''}, we analyze the remaining 45 responses and summarize them as follows. Note that a response may be summarized into multiple aspects.

\begin{itemize}
\item \textbf{Consistent to our attributes or evaluation criteria.} We find that the comments of 31 participants are consistent to our attributes or the evaluation criteria. For example, {\em ``Providing references (links to other discussions, code changes, documentation, etc.) is important for building trust and minimizing the length of feedback cycles''} is consistent to I.O2 and {\em ``Politeness''} is consistent to E.E2.
\item \textbf{Overall expectations on CRCs.} There are 11 participants comment their expectations on CRCs, which may not be directly related to clarity. For example, 
{\em ``The reviewer should be responsive i.e. should quickly respond to the questions raised by the contributor''} is about the time taken to reply and {\em ``Some reviews will require a history to develop how experienced the reviewee is so the appropriate level of explanation is given for their skill''} is about writing CRCs based on reviewee's knowledge.
\item \textbf{Other suggestions on code review.} There are 8 participants comment on other suggestions regarding the practice of code review. For example, 
{\em ``Rich features of the code review tool is also important. For example, GitHub provides a "suggestion" feature, it makes the suggestion of the code change clear''} is about leveraging tools to provide suggestions and {\em ``I'm not sure if this it too meta, but I think having a space to talk about what kind of culture you want to encourage is important to setting standards''} is about the role of code review in building relationship and culture.
\end{itemize}

Overall, there is no additional attribute derived from the practitioners' feedback on potential new attributes. However, they provide valuable insights of their expectations on code review, and may inspire future studies to improve the quality of CRCs and the practice of code review.

\greybox{Summary of RQ1}{Based on our literature review and survey with open-source practitioners, we derive three attributes related to the clarity of CRCs and their corresponding evaluation criteria. A majority of the practitioners consider these three attributes important to the clarity of CRCs.}

\subsection{\textbf{RQ2: Clarity of CRCs in Open-Source Projects}}
In this RQ, we first present our detailed process on analyzing the results of our manual investigation. We then present the results of our quantitative analysis and case study, respectively.

\phead{Experimental Setup.}
Two authors of this paper independently label the clarity of CRCs following the process and using the datasets discussed in Section~\ref{sec:methodology}.
Particularly, for each attribute, the CRC will be marked as {\em positive} if it meets {\em all} of the essential criteria and {\em at least one} of the optional criteria. Otherwise, it will be marked as {\em negative}. \change{During the data annotation, each attribute is separately and independently labelled, and the results for one indicator will not affect those of the other two attributes.}
When the labelling is completed, the two authors compare their results and discuss each disagreement until reaching a consensus. The value of Cohen's Kappa~\cite{kappa} in this process is 0.87, which indicates a substantial agreement.

\phead{Quantitative Analysis.} We present the results of our quantitative analysis on the clarity of CRCs by different programming languages.
Table~\ref{table:RQ2} shows the percentage of CRCs' clarity for each programming language. 
Specifically, \textit{``Negative''} refers to the percentage of CRCs that do not meet the evaluation criteria for each attribute, \textit{``All positive''} refers to the percentage of CRCs that meet the evaluation criteria for all the three attributes. 
Overall, 71.2\% of the CRCs meet the evaluation criteria in all of the three attributes, meaning that a large portion of the CRCs (i.e., 28.8\%) is not shown to have a sufficient clarity.
We discuss the results by comparing among the attributes and the programming languages, respectively.

\noindent \underline{Comparison among attributes.}
The distribution of CRCs that are negative for different attributes is 11.4\% on average for {\em Relevance}, 19.3\% on average for {\em Informativeness}, and 5.8\% on average for {\em Expression}, respectively. The results show that a non-negligible portion of CRCs in open-source projects is not written with good clarity, especially for {\em Informativeness} and {\em Relevance}.

\noindent \underline{Comparison among programming languages.}
We find that the distribution of clarity varies for different programming languages. For example, over 75\% of the CRCs for {\em C} and {\em Java} are all positive. Differently, only 63.6\% of the CRCs are all positive for {\em C++}, meaning that over 35\% of its CRCs have an insufficient clarity.

\phead{Case Study.} For each attribute, we discuss a negative example (i.e., does not meet the evaluation criteria discussed in the experimental setup of this RQ) and a positive example (i.e., meets the evaluation criteria), respectively. Note that we rename the identifier names and slightly rephrase the CRC to avoid directly retrieving the author of the CRC based on the examples in this paper.

\noindent \underline{Relevance.}
As presented in the examples below, the negative example comments {\em ``Same here. and also all others''}. The CRC itself hardly contains any information relevant to the code change. It may only be relevant to the information outside this code change. Therefore, it is not shown to be relevant to the code change, and it is not self-interpretable.
Therefore, this CRC does not meet the essential criteria of {\em R.E1}.
In comparison, the positive example raises a question, and the question is shown to be relevant to the code change. Moreover, it specifies the exact location in the code change where the reviewer has a question.

\begin{lstlisting}[language=Python, label=lst:r1, captionpos=t]
# Negative Example (Ruby)
+    def print_the_page(**options)
+      options[:page_ranges] &&= Array(options[:page_ranges])
+      bridge.print_the_page(options)
+    end

`\textcolor{red}{\textbf{CRC: Same here. and also all others.}}`

# Positive Example (Python)
-    self._internal = self._internal.resolved_copy
+    self._update_internal_frame(
+        self._internal.resolved_copy, requires_same_anchor=False

`\textcolor{blue}{\textbf{CRC: When do we need to set 'requires\_same\_anchor=False?'}}`
\end{lstlisting}

\noindent \underline{Informativeness.}
As shown in the negative example below, the CRC mentions ``\textit{This change is not correct}''. This CRC may imply the developer to revert this change or fix an issue. However, the comment does not provide an explanation of why the change is considered incorrect. It's important to offer specific reasons or context information to help the developer understand the issue.
Therefore, this example CRC does not meet the essential criteria of {\em I.E2}.
In comparison, the positive example explains the issue and further provides a suggestion for the further action to reproduce the problem.

\begin{lstlisting}[language=Python, label=lst:i1, captionpos=t]
# Negative Example (JavaScript)
       hosts.add(host);
-      LOG.info(String.format(""Added node %s."", node.getId()));
+      LOG.finest(String.format(""Added node %s."", node.getId()));
       host.runHealthCheck();

`\textcolor{red}{\textbf{CRC: This change is not correct.}}`

# Positive Example (JavaScript)
+       process.on('SIGUSR2', function () {
+               log.reopenFileStreams();
+       });
+
         module.exports.logger = logger;

`\textcolor{blue}{\textbf{CRC: Check the linting is failing, 'log' is not defined.}}`
`\textcolor{blue}{\textbf{You can run locally 'npm run lint' to double check.}}`
\end{lstlisting}

\noindent \underline{Expression.}
The CRC in the negative example asks the question in an impolite manner, which does not meet the essential evaluation criteria of {\em E.E2}. According to many of our survey participants' feedback, being polite and friendly is very important to efficient communications. Constructive criticism and polite suggestions for improvement are always preferred than harsh or toxic comments.

\begin{lstlisting}[language=Python, label=lst:e1, captionpos=t]
# Negative Example (C#)
-       (i1, s2, err2, s2) =>
+       (i1, s2, errCode, err2, s2) =>

`\textcolor{red}{\textbf{CRC: wtf is i1, s2, errCode, err2, s2?}}`
`\textcolor{red}{\textbf{I know we have nested lambdas by maybe this is a case for a method.}}`

# Positive Example (JavaScript)
 using Telemetry;
+using Telemetry.Trace;

`\textcolor{blue}{\textbf{CRC: This is technically correct, I wonder if we could make it}}`
`\textcolor{blue}{\textbf{simpler by not requiring this namespace?}}`
\end{lstlisting}

\greybox{Summary of RQ2}{We find that a large portion (i.e., 28.8\%) of the \change{CRCs} in our study open-source datasets have insufficient clarity. Among the three attributes, {\em Informativeness} has the most noticeable insufficiency.}

 \begin{table}
    \caption{ Distribution (\%) of CRCs' clarity for each programming language (RQ2).}
    \centering
    \resizebox{\columnwidth}{!} {
    \tabcolsep=20pt
    \begin{tabular}{l|c|c|c|c}

\hline
\multirow{2}{*}{\textbf{Language}} & \multicolumn{3}{c|}{\textbf{Negative}}   &  \cellcolor[gray]{0.90}\\ \cline{2-4}
    & Relevance & Informativeness & Expression   &  \multirow{-2}{*}{\textbf{All Positive}} \cellcolor[gray]{0.90}\\
\hline

        \textbf{C} & 11.1            & 14.4         & 5.1          &  77.3   \\
        \textbf{C++}  & 11.5              & 28.1          & 6.3            & 63.6     \\
        \textbf{C\#} & 9.8             & 25.6          & 8.9            & 63.8     \\
        \textbf{Golang}  & 11.2              & 16.2          &  3.8            & 74.0      \\
        \textbf{Java}  & 6.7              & 17.9          & 6.1             & 75.3      \\
        \textbf{JavaScript} & 9.3              & 17.1        & 5.3             & 72.2    \\
        \textbf{PHP}  & 15.0              & 15.5      & 3.4             & 70.4      \\
        \textbf{Python}   &  11.2              & 18.2      & 4.0            & 73.6     \\
        \textbf{Ruby} & 18.4              & 20.9      & 9.6            & 68.8    \\

        \hline
        \textit{Overall}     & 11.4          & 19.3  & 5.8   & 71.2    \\

        \hline
    \end{tabular}
    }
    \vspace{-0.0cm}

    \label{table:RQ2}
\end{table}

\subsection{\textbf{RQ3: Automatically Evaluating the Clarity of CRCs}}
In this RQ, we present the results of our evaluation on the clarity of CRCs.

\subsubsection{\change{Main Results.}}
\noindent \change{Table~\ref{table:RQ3} presents the results of ClearCRC and it is organized by different \textsf{RIE} attributes, model, and metrics. In the table, we report the average results of five-fold cross validation. The bold number of each column shows the best performance of the corresponding attribute and metric.} 
\mbox{}

\noindent \underline{\change{Comparison among model sets.}}
\change{Overall, pre-trained language models achieve the best performance among all model sets. 
Specifically, with regard to the set average performance across all evaluation attributes, pre-trained language models have the best performance on all four metrics. For example, the balanced accuracy of set 2 models is 71.25\%, outperforming other sets by a large margin (i.e., 17.7\% and 20.8\% relative improvements compared to set 1 and set 3 models, respectively). Apart from balanced accuracy, there is also a consistent advantage for set 2 models on the other three metrics. We believe that the strong performance of pre-trained language models is attributed to their prior code knowledge and task-specific fine-tuning. In contrast, large language models perform poorly due to their inability to acquire sufficient knowledge about the clarity of CRCs.}

\noindent \underline{\change{Comparison among attributes.}} \change{According to our experimental results, we find that the performance on different attributes varies a lot. The average balanced accuracy of \textit{Informativeness} is 71.30\%, while the number is 56.91\% for \textit{Expression}, which may suggest that \textit{Expression} is a relatively easier attribute for subject models to understand and recognize, but harder for \textit{Informativeness}. However, we also notice that the recall of \textit{Informativeness} is much lower than \textit{Relevance} and \textit{Expression} (i.e., 78.22\% compared to 83.41\% and 95.87\%). We conjecture that since the dataset of \textit{Informativeness} has a higher proportion of negative samples, the model achieves a higher balanced accuracy in distinguishing positive and negative samples while also making it more difficult to recall negative samples.}

\noindent \underline{\change{Comparison within the set.}} 
\change{We mainly analyze the results in set 2 models because they are the best set among all models. For average performance among all attributes, CodeBERT is better in terms of balanced accuracy (i.e., 73.04\% > 69.46\%) but behind CodeReviewer on F-1 score (i.e., 93.07\% > 94.61\%). 
Overall, the two models have comparable strengths. Considering that the size of CodeBERT (i.e., 125M) is about half of CodeReviewer's (i.e., 223M), we believe CodeBERT demonstrates good generalizability for automatically evaluating CRC's clarity.}

\change{Overall, the results show that in terms of precision, eecall, and F-1 score, the performance of these models is satisfactory (i.e., an overall average score of more than 85\%), but there is still room for improvement for balanced accuracy (i.e., 63.58\%), which could be credited to the suboptimal ability to detect negative CRC samples. Therefore, we believe that ClearCRC is promising in automatically evaluating the clarity of CRCs and further exploration is still required.}

\subsubsection{\change{Generalizability on other datasets.}}
\change{To evaluate if \tool could generalize to newer or less-studied projects, we conduct a study on a subset of CodeReviewer-New~\cite{guo2024exploring}, which includes repositories that the original CodeReviewer dataset does not contain, and adopts various approaches to ensure the data quality. We randomly sample 135 examples from CodeReviewer-New (i.e., 15 samples for each of the 9 languages), and follow the same data annotation approaches and experimental settings as the main experiments. We use the best checkpoint in the fold 1 cross validation of the main experiments for each model.}

\change{Compared to the results with the original CodeReviewer dataset, there is a slight drop of the performance for both models. For example, the balanced ccuracy decreases by around 3\% (i.e., 71.25\% -> 68.14\%) and the F-1 score decreases by 5\% (i.e., 93.84\% -> 88.08\%). Considering the different time and project distributions of the two datasets, we think that the decline is still reasonable and \tool could be generalized to the additional datasets.
}

\greybox{Summary of RQ3}{
\change{\tool assisted with pre-trained language models shows promising results for automatic evaluation of the clarity of CRCs, with a balanced accuracy and F-1 score of 71.25\% and 93.84\%, respectively. Additionally, it could be generalized to newer or less-studied datasets.}
}

\begin{table}
    \caption{ \change{Balanced Accuracy (BA, \%), Precision (P, \%), Recall (R, \%), and F-1 score (\%) of \tool using different models. \textbf{Bold} value: Maximum value in each metric of each attribute.} }
    %\scriptsize
    \centering
    \resizebox{\columnwidth}{!} {
    \tabcolsep=2.5
    pt
    \renewcommand\arraystretch{1.15}
    
\begin{tabular}{|ccccccccccccccccc|}
\hline
\multicolumn{1}{|c|}{\multirow{2}{*}{\textbf{Models}}} & \multicolumn{4}{c|}{\textbf{Relevance}} & \multicolumn{4}{c|}{\textbf{Informativeness}} & \multicolumn{4}{c|}{\textbf{Expression}} & \multicolumn{4}{c|}{\textbf{Average}} \\ \cline{2-17} 
\multicolumn{1}{|c|}{} & \multicolumn{1}{c|}{\textbf{BA}} & \multicolumn{1}{c|}{\textbf{P}} & \multicolumn{1}{c|}{\textbf{R}} & \multicolumn{1}{c|}{\textbf{F1}} & \multicolumn{1}{c|}{\textbf{BA}} & \multicolumn{1}{c|}{\textbf{P}} & \multicolumn{1}{c|}{\textbf{R}} & \multicolumn{1}{c|}{\textbf{F1}} & \multicolumn{1}{c|}{\textbf{BA}} & \multicolumn{1}{c|}{\textbf{P}} & \multicolumn{1}{c|}{\textbf{R}} & \multicolumn{1}{c|}{\textbf{F1}} & \multicolumn{1}{c|}{\textbf{BA}} & \multicolumn{1}{c|}{\textbf{P}} & \multicolumn{1}{c|}{\textbf{R}} & \textbf{F1} \\ \hline
\multicolumn{17}{|c|}{\textit{Set 1: Machine Learning and Deep Learing Models}} \\ \hline
\multicolumn{1}{|c|}{\textbf{Random Forest}} & \multicolumn{1}{c|}{52.77} & \multicolumn{1}{c|}{88.04} & \multicolumn{1}{c|}{\textbf{99.91}} & \multicolumn{1}{c|}{93.60} & \multicolumn{1}{c|}{69.90} & \multicolumn{1}{c|}{89.19} & \multicolumn{1}{c|}{87.60} & \multicolumn{1}{c|}{88.38} & \multicolumn{1}{c|}{53.13} & \multicolumn{1}{c|}{94.73} & \multicolumn{1}{c|}{\textbf{99.82}} & \multicolumn{1}{c|}{97.21} & \multicolumn{1}{c|}{58.60} & \multicolumn{1}{c|}{90.66} & \multicolumn{1}{c|}{95.78} & 93.06 \\ \hline
\multicolumn{1}{|c|}{\textbf{LSTM}} & \multicolumn{1}{c|}{58.66} & \multicolumn{1}{c|}{89.58} & \multicolumn{1}{c|}{95.50} & \multicolumn{1}{c|}{92.39} & \multicolumn{1}{c|}{68.93} & \multicolumn{1}{c|}{90.70} & \multicolumn{1}{c|}{72.49} & \multicolumn{1}{c|}{80.19} & \multicolumn{1}{c|}{59.78} & \multicolumn{1}{c|}{95.48} & \multicolumn{1}{c|}{96.72} & \multicolumn{1}{c|}{96.08} & \multicolumn{1}{c|}{62.45} & \multicolumn{1}{c|}{91.92} & \multicolumn{1}{c|}{88.24} & 89.55 \\ \hline
\multicolumn{1}{|c|}{\textit{Set 1 Avg.}} & \multicolumn{1}{c|}{\textit{55.72}} & \multicolumn{1}{c|}{\textit{88.81}} & \multicolumn{1}{c|}{\textit{97.70}} & \multicolumn{1}{c|}{\textit{92.99}} & \multicolumn{1}{c|}{\textit{69.41}} & \multicolumn{1}{c|}{\textit{89.95}} & \multicolumn{1}{c|}{\textit{80.04}} & \multicolumn{1}{c|}{\textit{84.28}} & \multicolumn{1}{c|}{\textit{56.45}} & \multicolumn{1}{c|}{\textit{95.11}} & \multicolumn{1}{c|}{\textit{98.27}} & \multicolumn{1}{c|}{\textit{96.64}} & \multicolumn{1}{c|}{\textit{60.53}} & \multicolumn{1}{c|}{\textit{91.29}} & \multicolumn{1}{c|}{\textit{92.01}} & \textit{91.31} \\ \hline
\multicolumn{17}{|c|}{\textit{Set 2: Pre-trained Language Models}} \\ \hline
\multicolumn{1}{|c|}{\textbf{CodeBERT}} & \multicolumn{1}{c|}{\textbf{78.37}} & \multicolumn{1}{c|}{\textbf{95.12}} & \multicolumn{1}{c|}{88.48} & \multicolumn{1}{c|}{91.27} & \multicolumn{1}{c|}{77.43} & \multicolumn{1}{c|}{92.22} & \multicolumn{1}{c|}{88.78} & \multicolumn{1}{c|}{90.42} & \multicolumn{1}{c|}{\textbf{63.31}} & \multicolumn{1}{c|}{\textbf{95.89}} & \multicolumn{1}{c|}{99.22} & \multicolumn{1}{c|}{\textbf{97.53}} & \multicolumn{1}{c|}{\textbf{73.04}} & \multicolumn{1}{c|}{\textbf{94.41}} & \multicolumn{1}{c|}{92.16} & 93.07 \\ \hline
\multicolumn{1}{|c|}{\textbf{CodeReviewer}} & \multicolumn{1}{c|}{76.83} & \multicolumn{1}{c|}{93.91} & \multicolumn{1}{c|}{97.30} & \multicolumn{1}{c|}{\textbf{95.55}} & \multicolumn{1}{c|}{73.02} & \multicolumn{1}{c|}{90.02} & \multicolumn{1}{c|}{\textbf{92.95}} & \multicolumn{1}{c|}{\textbf{91.41}} & \multicolumn{1}{c|}{58.53} & \multicolumn{1}{c|}{95.39} & \multicolumn{1}{c|}{98.44} & \multicolumn{1}{c|}{96.87} & \multicolumn{1}{c|}{69.46} & \multicolumn{1}{c|}{93.11} & \multicolumn{1}{c|}{\textbf{96.23}} & \textbf{94.61} \\ \hline
\multicolumn{1}{|c|}{\textit{Set 2 Avg.}} & \multicolumn{1}{c|}{\textit{77.60}} & \multicolumn{1}{c|}{\textit{94.51}} & \multicolumn{1}{c|}{\textit{92.89}} & \multicolumn{1}{c|}{\textit{93.41}} & \multicolumn{1}{c|}{\textit{75.23}} & \multicolumn{1}{c|}{\textit{91.12}} & \multicolumn{1}{c|}{\textit{90.86}} & \multicolumn{1}{c|}{\textit{90.92}} & \multicolumn{1}{c|}{\textit{60.92}} & \multicolumn{1}{c|}{\textit{95.64}} & \multicolumn{1}{c|}{\textit{98.83}} & \multicolumn{1}{c|}{\textit{97.20}} & \multicolumn{1}{c|}{\textit{71.25}} & \multicolumn{1}{c|}{\textit{93.76}} & \multicolumn{1}{c|}{\textit{94.19}} & \textit{93.84} \\ \hline
\multicolumn{17}{|c|}{\textit{Set 3: Large Language Models}} \\ \hline
\multicolumn{1}{|c|}{\textbf{Llama}} & \multicolumn{1}{c|}{61.76} & \multicolumn{1}{c|}{90.42} & \multicolumn{1}{c|}{86.90} & \multicolumn{1}{c|}{88.60} & \multicolumn{1}{c|}{\textbf{78.62}} & \multicolumn{1}{c|}{\textbf{94.12}} & \multicolumn{1}{c|}{79.68} & \multicolumn{1}{c|}{86.30} & \multicolumn{1}{c|}{52.14} & \multicolumn{1}{c|}{94.58} & \multicolumn{1}{c|}{98.16} & \multicolumn{1}{c|}{96.90} & \multicolumn{1}{c|}{64.17} & \multicolumn{1}{c|}{93.04} & \multicolumn{1}{c|}{88.25} & 90.60 \\ \hline
\multicolumn{1}{|c|}{\textbf{CodeLlama}} & \multicolumn{1}{c|}{46.74} & \multicolumn{1}{c|}{85.16} & \multicolumn{1}{c|}{32.36} & \multicolumn{1}{c|}{46.90} & \multicolumn{1}{c|}{59.92} & \multicolumn{1}{c|}{88.72} & \multicolumn{1}{c|}{47.82} & \multicolumn{1}{c|}{62.12} & \multicolumn{1}{c|}{54.56} & \multicolumn{1}{c|}{94.94} & \multicolumn{1}{c|}{82.84} & \multicolumn{1}{c|}{88.46} & \multicolumn{1}{c|}{53.74} & \multicolumn{1}{c|}{89.61} & \multicolumn{1}{c|}{54.34} & 65.83 \\ \hline
\multicolumn{1}{|c|}{\textit{Set 3 Avg.}} & \multicolumn{1}{c|}{54.25} & \multicolumn{1}{c|}{87.79} & \multicolumn{1}{c|}{59.63} & \multicolumn{1}{c|}{67.75} & \multicolumn{1}{c|}{69.27} & \multicolumn{1}{c|}{91.42} & \multicolumn{1}{c|}{63.75} & \multicolumn{1}{c|}{74.21} & \multicolumn{1}{c|}{53.35} & \multicolumn{1}{c|}{94.76} & \multicolumn{1}{c|}{90.50} & \multicolumn{1}{c|}{92.68} & \multicolumn{1}{c|}{58.96} & \multicolumn{1}{c|}{91.32} & \multicolumn{1}{c|}{71.29} & 78.21 \\ \hline
\multicolumn{17}{|c|}{\textit{Overall}} \\ \hline
\multicolumn{1}{|c|}{\textbf{Average}} & \multicolumn{1}{c|}{62.52} & \multicolumn{1}{c|}{90.37} & \multicolumn{1}{c|}{83.41} & \multicolumn{1}{c|}{84.72} & \multicolumn{1}{c|}{71.30} & \multicolumn{1}{c|}{90.83} & \multicolumn{1}{c|}{78.22} & \multicolumn{1}{c|}{83.14} & \multicolumn{1}{c|}{56.91} & \multicolumn{1}{c|}{95.17} & \multicolumn{1}{c|}{95.87} & \multicolumn{1}{c|}{95.51} & \multicolumn{1}{c|}{63.58} & \multicolumn{1}{c|}{92.12} & \multicolumn{1}{c|}{85.83} & 87.79 \\ \hline
\end{tabular}
}
\vspace{-0.0cm}
\label{table:RQ3}
\end{table}

\begin{table}
    \caption{ \change{Results on CodeReviewer-New for set 2 models.}}
    \centering
    \resizebox{\columnwidth}{!} {
    \tabcolsep=2.5
    pt
    \renewcommand\arraystretch{1.15}
    
\begin{tabular}{|c|cccc|cccc|cccc|cccc|}
\hline
\multirow{2}{*}{\textbf{Models}} & \multicolumn{4}{c|}{\textbf{Relevance}} & \multicolumn{4}{c|}{\textbf{Informativeness}} & \multicolumn{4}{c|}{\textbf{Expression}} & \multicolumn{4}{c|}{\textbf{Average}} \\ \cline{2-17} 
 & \multicolumn{1}{c|}{\textbf{BA}} & \multicolumn{1}{c|}{\textbf{P}} & \multicolumn{1}{c|}{\textbf{R}} & \textbf{F1} & \multicolumn{1}{c|}{\textbf{BA}} & \multicolumn{1}{c|}{\textbf{P}} & \multicolumn{1}{c|}{\textbf{R}} & \textbf{F1} & \multicolumn{1}{c|}{\textbf{BA}} & \multicolumn{1}{c|}{\textbf{P}} & \multicolumn{1}{c|}{\textbf{R}} & \textbf{F1} & \multicolumn{1}{c|}{\textbf{BA}} & \multicolumn{1}{c|}{\textbf{P}} & \multicolumn{1}{c|}{\textbf{R}} & \textbf{F1} \\ \hline
\textbf{CodeBERT} & \multicolumn{1}{c|}{61.39} & \multicolumn{1}{c|}{91.94} & \multicolumn{1}{c|}{94.21} & 93.06 & \multicolumn{1}{c|}{63.27} & \multicolumn{1}{c|}{74.77} & \multicolumn{1}{c|}{87.91} & 80.81 & \multicolumn{1}{c|}{75.45} & \multicolumn{1}{c|}{88.12} & \multicolumn{1}{c|}{87.25} & 87.68 & \multicolumn{1}{c|}{66.70} & \multicolumn{1}{c|}{84.94} & \multicolumn{1}{c|}{89.79} & 87.18 \\ \hline
\textbf{CodeReviewer} & \multicolumn{1}{c|}{64.55} & \multicolumn{1}{c|}{92.62} & \multicolumn{1}{c|}{93.39} & 93.00 & \multicolumn{1}{c|}{63.67} & \multicolumn{1}{c|}{74.17} & \multicolumn{1}{c|}{97.80} & 84.36 & \multicolumn{1}{c|}{80.48} & \multicolumn{1}{c|}{90.91} & \multicolumn{1}{c|}{88.24} & 89.55 & \multicolumn{1}{c|}{69.57} & \multicolumn{1}{c|}{85.90} & \multicolumn{1}{c|}{93.14} & 88.97 \\ \hline
\textbf{Average} & \multicolumn{1}{c|}{62.97} & \multicolumn{1}{c|}{92.28} & \multicolumn{1}{c|}{93.80} & 93.03 & \multicolumn{1}{c|}{63.47} & \multicolumn{1}{c|}{74.47} & \multicolumn{1}{c|}{92.86} & 82.59 & \multicolumn{1}{c|}{77.97} & \multicolumn{1}{c|}{89.52} & \multicolumn{1}{c|}{87.75} & 88.62 & \multicolumn{1}{c|}{68.14} & \multicolumn{1}{c|}{85.42} & \multicolumn{1}{c|}{91.47} & 88.08 \\ \hline
\end{tabular}
}
\vspace{-0.1cm}
\label{table:RQ3_2}
\end{table}

\vspace{-0.2cm}
\section{Discussion}
\label{sec:discussion}

\subsection{Implications}

\phead{Implication 1: Actionable guidelines for evaluating and writing clear CRCs.}
As shown in our results of RQ2, there is a large portion of the CRCs in open-source systems (i.e., 28.8\%) that lack clarity.
Due to the lack of well-defined guidelines on writing CRCs, it is challenging for developers to write CRCs that can clearly and sufficiently serve as the medium between developers and reviewers. Based on our survey with open-source practitioners, many participants mention that they do not want to see CRCs that are {\em ``confusing''} and {\em ``vague''}. Instead, they expect the CRCs to be {\em ``clear''}. However, it is also difficult to determine what is a {\em ``clear''} CRC. 

In our study, we characterize the clarity of CRCs and derive three attributes with their corresponding evaluation criteria. One of our survey participants mentions 
{\em ``I like the idea of thinking more about comments. I would welcome good guidelines.''}.
Therefore, our findings can be used as actionable guidelines for evaluating and writing clear CRCs, which in turn improves the efficiency and quality of the code review process.

\phead{Implication 2: Select high-quality data for the automated generation of CRCs.}
There are a series of studies that utilize existing CRC data to train models for the automated generation of CRCs~\cite{codereviewer, lu2023llama, tufano2022using}. However, these studies accept all the CRC data in general, without a curation or selection on the quality of data. Based on such a situation, existing CRC generation techniques may learn from CRC data with insufficient clarity and then generate confusing results. 

\begin{figure}
 \centering
\includegraphics[width=0.5\linewidth]{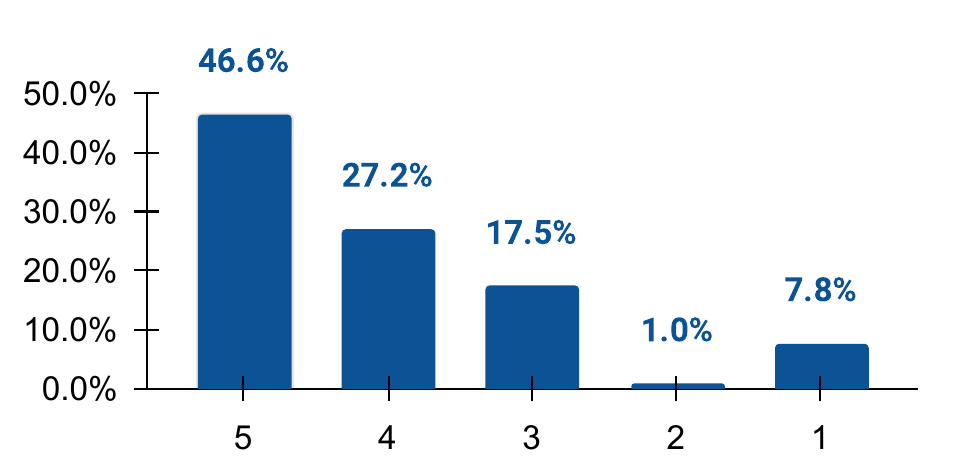}
 \caption{Rating of importance for the clarity of generated CRC in our survey.}
 \label{fig:clarity_rate_CRCG}
 \vspace{-0.0cm}
 \end{figure}

In our survey, we also ask the participants for their opinion on automated CRC generation techniques. They can rate the importance of the clarity of automatically generated CRCs from 1 to 5, where 1 indicates the lowest importance and 5 indicates the hightest importance. 
Figure~\ref{fig:clarity_rate_CRCG} presents the results of their ratings. The average rating is 4.04, and more than 70\% of the participants consider its importance as 4 or 5. For example, one participant comments that {\em ``If I am going to receive automated comments on my code changes, they need to be clear, accurate, and relevant. Otherwise they are just wasting my time''}.
As shown in the results of RQ3, \tool achieves a high precision in evaluating the clarity of CRCs on all the three attributes. Therefore, the findings of our study can be used for implementing data filtering and selection mechanisms to help identify CRCs with good clarity, improving the overall effectiveness of the automated CRC generation process.

\phead{Implication 3: Provide a more comprehensive quality evaluation for CRCs and its generation.}
As discussed above, developers expect to have CRCs with good quality to foster an effective communication among the team members. Moreover, existing research on automated CRC generation generally uses BLEU score~\cite{bleu} to examine the textual similarity between the generated CRC and the reference CRC. 
While BLEU score can be used to assess the performance of automated CRC generation techniques, it does not directly address the quality of the CRC itself. In other words, a high Bleu score does not necessarily indicate that the generated CRC is clear and concise.
Therefore, the quality of the CRC itself still remains unclear.

In this paper, we study the quality of CRC by understanding and uncovering the clarity of CRC. 
To do so, we derive the {\sf RIE} attributes (i.e., Relevance, Informativeness, and Expressiveness) and their respective evaluation criteria. These attributes and criteria can be leveraged to provide a more comprehensive evaluation for CRCs and their automated generation.
By utilizing our findings, we aim to provide a more comprehensive quality evaluation for CRCs and their automated generation. It can help developers in writing better CRCs, and also contribute to the improvement of automated CRC generation techniques, ultimately leading to more effective communication among software development teams.

\subsection{\change{\textsf{RIE} Indicators and Existing Metrics}}

\phead{\change{Comparison with Existing Metrics.}}
\change{Despite the \textsf{RIE} attributes we derive from this paper, the community and research have proposed other metrics to evaluate the quality of CRCs such as Readability, Sentiment~\cite{ASRI201937}, and Usefulness~\cite{7962371}. Below, we compare these existing metrics with our \textsf{RIE} attributes:}

\begin{itemize}
    \item \change{\textbf{Readability}: Readability is seen as an important quality dimension of software comments~\cite{eleyan2020enhancing}, and it emphasizes the difficulty of reading the text (e.g., the number of difficult words and length of the sentence). It is a subset of our Expression indicator (i.e., E.O1), and Expression includes other aspects like the tone. Besides, a readable CRC could easily be unclear. For example, a comment generated by LLM is typically very easy to read, but it may contain little valuable and helpful information for further action.} 

    \item \change{\textbf{Sentiment}~\cite{ASRI201937}: In code review activities, the contributors frequently express positive, neutral, and negative sentiments, and these sentiments are correlated with the complete time of review~\cite{ASRI201937}. Sentiment is related to the evaluation criteria of Expression (i.e., E.E2 ``Polite and objective''), as polite and objective comments are usually positive or neutral. However, sentiment alone is not enough to assess the clarity of CRC---purely encouraging and positive comments are not necessarily clear, while neutral CRC could achieve sufficient clarity.}

    \item \change{\textbf{Usefulness}~\cite{7962371}: Usefulness of one CRC is typically measured based on the outcome of the corresponding code change (e.g., acceptance rate and time) in former research~\cite{7962371}, on the assumption that more and faster code changes are useful because they can lead to better software quality. However, the outcome of one code change depends on many other factors like the identity of the contributor, code style, and the change scope~\cite{ram2018makes}, which adds more indeterministic to this measure. In contrast to Usefulness, which could be deemed as an outcome-driven metric, the \textsf{RIE} attributes are driven by the process of code review activities, mainly focusing on the clarity of CRCs themselves.}
\end{itemize}

\noindent \change{In this paper, instead of introducing a single new metric, we present a set of well-defined and actionable evaluation criteria for assessing the clarity of CRCs. These criteria are derived from a systematic process and align with the shared expectations of practitioners across academia, industry, and the open-source community.}

\phead{\change{Relation Among \textsf{RIE} attributes.}}
 \change{The \textsf{RIE} attributes aim to measure the clarity of CRCs from three distinct dimensions. They are conceptually orthogonal, meaning that each dimension assesses a distinct quality aspect and is relatively independent of the others:}
 
\begin{itemize}
    \item \change{\textbf{Relevance}: \textit{Relevance} primarily assesses the degree and correctness of the relevance between CRCs and code changes.}
    \item \change{\textbf{Informativeness}: \textit{Informativeness} mainly evaluates whether one CRC provides useful information like intention, explanation, suggestion, and reference information.}
    \item \change{\textbf{Expression}: \textit{Expression} measures if the CRC is expressed appropriately from perspectives like readability.}
\end{itemize}

\noindent \change{Since these dimensions capture different aspects, a CRC may perform well in one dimension but poorly in another. For example, a CRC may be highly relevant to code but lack meaningful information, or it may contain rich details but be poorly written and hard to understand. Because of this, the three dimensions should be evaluated separately to comprehensively assess the clarity of CRCs. 
Besides, we would like to mention that although the concepts and evaluation criteria are independent, some attributes are indeed correlated and likely co-occur. For example, if one code reviewer is merely complaining about a specific code change, the comment is typically neither informative nor polite.}
\section{Threats to Validity}
\label{sec:threats}

\phead{Internal Validity.}
We manually label the clarity of CRCs on open-source datasets. To mitigate the subjective bias in this process, two authors of this paper label the data independently. The labellers then discuss each disagreement until a consensus is reached. 
Following prior studies~\cite{ding2023temporal,li2023icse,wheretolog_ASE20}, we use Cohen's Kappa~\cite{kappa} to measure the agreement of the manual investigation results between the two authors.
The Cohen’s Kappa value in this process is 0.87, which indicates a substantial agreement.
The randomness in the process of our experiments (e.g., splitting the data, training the models) may affect the results.
To mitigate such threats, we use a five-fold cross validation to conduct the experiments and report the average number in our discussions.
We derive the evaluation criteria by analyzing the 103 survey responses from participants. Engaging more experts from various domains may generate more comprehensive results.

\phead{External Validity.}
We conduct our study on the datasets proposed by Li et al.~\cite{codereviewer}. Using other datasets may generate different results and findings. However, the datasets of Li et al.~\cite{codereviewer} extract code changes and the corresponding CRCs from open-source projects written in nine programming languages, which include a diverse range of repositories.
We derive the detailed evaluation criteria based on the survey with practitioners from open-source projects written in nine programming languages.  However, the findings of our study are not specialized for specific programming languages and can be generalizable to various projects. 
Our study is conducted based on open-source data and practitioners. Future studies may verify the generalizability of our findings on industrial systems and projects.
 \vspace{-0.1cm}
\section{Conclusion}
 \vspace{-0.1cm}
\label{sec:conclusion}

In this paper, we investigate how a code review comment (CRC) can clearly and concisely serve as the medium of communication among developers by conducting a multi-phased, comprehensive study. 
We derive our \textsf{RIE} attributes of the clarity of CRCs and the detailed evaluation criteria based on the analysis of our literature review and survey with practitioners. 
We also find that a noticeable portion of the CRCs in open-source projects do not have sufficient clarity. 
We further seek to explore the potential of automatically evaluating the clarity of CRCs by proposing an automated framework, namely \tool. Experimental results show that \tool is effective in evaluating the clarity of CRCs based on our {\sf RIE} attributes and outperforms the baseline approaches by a considerable margin. Our findings shed light on characterizing the quality of CRCs and further facilitate the collaboration between developers.

\section*{Data Availability}
Our replication package is available and can be accessed using the link~\cite{replication_package}.

\section*{Acknowledgment}
This work was supported by the National Key R\&D Program of China (No. 2024YFB4506400), sponsored by CCF-Huawei Populus Grove Fund, and National Natural Science Foundation of China (No. 62172214).

\balance
\bibliographystyle{ACM-Reference-Format}

\bibliography{paper}

\end{document}